\documentclass[11pt,a4paper]{article}
\usepackage[utf8]{inputenc}
\usepackage{amsmath}
\usepackage{amsfonts}
\usepackage{amssymb}
\usepackage{amsbsy}
\usepackage{graphicx}
\usepackage{caption}

\usepackage{threeparttable}
\usepackage{changepage}
\usepackage{placeins}

\usepackage{tikz}
\usetikzlibrary{positioning,fit,arrows,shapes.geometric,chains,decorations.pathreplacing,arrows.meta}

\usepackage[super]{nth}

\usepackage{hyperref}
\hypersetup{pdfborder={0 0 1}}
\usepackage{adjustbox}

\usepackage[left=3cm,right=3cm,top=2.5cm,bottom=2.5cm]{geometry}
\usepackage{mathrsfs}

\usepackage{tikz}

\usepackage[american]{babel}
\usepackage{csquotes}

\usepackage[%
    style=numeric,
    natbib=true,
    giveninits=true,
    uniquename=init,
    alldates=year,
    backend=biber]{biblatex}
\DeclareLanguageMapping{american}{american-apa}

\newrobustcmd*{\parentexttrack}[1]{%
  \begingroup
  \blx@blxinit
  \blx@setsfcodes
  \blx@bibopenparen#1\blx@bibcloseparen
  \endgroup}

\title{\textbf{Predicting Status of Pre and Post M\&A Deals Using Machine Learning and Deep Learning Techniques\thanks{
We are very grateful to Satyan Malhotra, CEO at Ask2.ai, Inc as the Senior Advisor and Industry Expert. We would like to thank 
Zhenyi Cao, Sai Varun Challa, Sahil	Chutani, Xuesong Ding, Junying Fang, Yuheng Han, Ruien Jiang, Jean Kergall, Sumeet Kotaria, Rohit	Kumar, Junru Liu, Jeff	Lv, Cyrus Moazami, Minxuan	Nie, Gaolin	Qian, Neel	Save, Xinyue Tian, Xiao	Yang for their participation and help on this research. Errors are our own responsibility.}}}
\date{}
\author{Tugce Karatas\footnote{Department of IEOR, Columbia University, \textbf{tk2757@columbia.edu}} \and  Ali Hirsa\footnote{Department of IEOR, Columbia University, \textbf{ah2347@columbia.edu}}}

\begin{document}

\maketitle
	
\begin{abstract}
Risk arbitrage or merger arbitrage is a well-known investment strategy that speculates on the success of M\&A deals. Prediction of the deal status in advance is of great importance for risk arbitrageurs. If a deal is mistakenly classified as a completed deal, then enormous cost can be incurred as a result of investing in target company shares. On the contrary, risk arbitrageurs may lose the opportunity of making profit.  In this paper, we present an \texttt{ML} and \texttt{DL} based methodology for takeover success prediction problem. We initially apply various \texttt{ML} techniques for data preprocessing such as \texttt{kNN} for data imputation, \texttt{PCA} for lower dimensional representation of numerical variables, \texttt{MCA} for categorical variables, and \texttt{LSTM} autoencoder for sentiment scores. We experiment with different cost functions, different evaluation metrics, and oversampling techniques to address class imbalance in our dataset. We then implement feedforward neural networks to predict the success of the deal status. Our preliminary results indicate that our methodology outperforms the benchmark models such as logit and weighted logit models. We also integrate sentiment scores into our methodology using different model architectures, but our preliminary results show that the performance is not changing much compared to the simple \texttt{FFNN} framework. We will explore different architectures and employ a thorough hyperparameter tuning for sentiment scores as a future work.
\end{abstract}

\providecommand{\keywords}[1]{\textbf{\textit{Keywords:}} #1}
\keywords{M\&A, takeover success, neural networks, PCA, MCA, kNN, LSTM autoencoder, SMOTE}

\section{Introduction}

When a company is willing to acquire another company, it agrees to pay a premium over the unaffected share price of the target company. Typically, the share price of the target company runs up close to the acquisition price on the day of the deal announcement. Once the deal is announced, the share price will reach a level less than the acquisition price as there is an uncertainty on the success of the deal. If the market believes that this deal will be completed successfully, usually post-announcement price will be very close to the acquisition price. On the other hand, if the success of the deal is suspicious, there will be a clear difference between acquisition price and the post-announcement price. Furthermore, there could be rumors on the merger and acquistions deals before the announcement date. These rumors are either released from the inside of the companies or the changes in the shares prices of the companies may trigger these rumors. Depending on the market's belief on the rumor, there is a possible increase in the share price of the target company. Therefore, predicting if there will be an announcement related to the deal and predicting if the deal will be successful after the announcement are of great importance as these predictions are very useful for the investment strategy that speculates on the successful completion of mergers and acquisitions. This strategy is known as \textit{risk arbitrage}, or \textit{merger arbitrage}. In this context, the arbitrageur capitalizes on the spread between the share price of the company and the acquisition price. If the arbitrageur believes that the deal will be completed successfully, he/she takes a long position in the share price of the target company and a short position in the share price of the acquirer company. Otherwise, he/she may take a short position in target company's shares and a long position in acquirer's shares. Once the deal is successfully completed, the profit of the arbitrageur will depend on the arbitrage spread at the time of the purchase. 
\begin{figure}[!htbp]
    \centering
    \includegraphics[scale = 0.5]{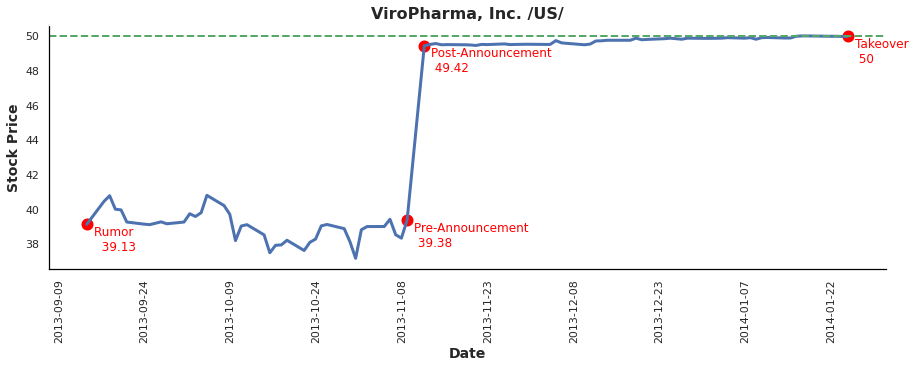}
    \caption{A Real Life Example of Stock Price Movement for Target Company during M\&A Period}
    \label{fig:example_takeover}
\end{figure}

Figure \ref{fig:example_takeover} shows a real life example of the merger and acquisition process. In September 2013, a rumor regarding the purchase of ViroPharma Incorporated by Irish drug manufacturer Shire plc surfaced. The deal announcement was made on November 11, 2013 with the agreed price of \$50 per share. Although ViroPharma's closing price was \$39.38 per share on the last trading day before the announcement, it suddenly jumped to \$49.42 on the day of the announcement. The deal was completed successfully on January 24, 2014 with the share price of \$50 as expected. The price movement of the share price of the target company in this example clearly shows that investment strategies can be built on using prediction of the deal announcement and then prediction of the takeover success in advance.

In this paper, we address two related questions. Figure \ref{plot:ma_framework} gives an overall framework of the process. In the first phase, we will work on predicting if there will be a deal announcement given that there are rumors on the deal. In the second phase, we will build a classification model for the success of the deal after the announcement. In both phases, we will be working with highly imbalanced data sets. Therefore, we will start by applying data preprocessing techniques on our data, and then propose a prediction and classification model for both phases. There are several studies on the prediction of success of takeover bids in the literature, but the focus has been only on a small portion of the problem. We will propose a more thorough framework in this study. To the best of our knowledge, prediction of the deal announcement based on the given rumor has not been considered in the literature. 
\begin{center}
\begin{tikzpicture}[auto, thick, >=triangle 45]
\matrix [column sep={25mm}, row sep=2.5mm] {
    & & \node(n1) [draw, shape=rectangle,rounded corners,fill=blue!20, minimum height=1.5em]{\textbf{Deal Complete}};
    \\
    & \node(n2)  [draw, shape=rectangle,rounded corners,fill=blue!20, minimum height=1.5em]{\textbf{Deal Announced}}; & \\
    & &
    \node(n4) [draw, shape=rectangle,rounded corners,fill=blue!20, minimum height=1.5em] {\textbf{Deal Cancelled}};\\
    \node(n3) [draw, shape=rectangle,rounded corners,fill=blue!20, minimum height=1.5em] {\textbf{Rumor}};& &\\
    \\
    & \node(n5) [draw, shape=rectangle,rounded corners,fill=blue!20, minimum height=1.5em] {\textbf{Rumor Cancelled}};& \\
};

 \tikzset{blue dotted/.style={draw=blue, line width=1pt,dash pattern=on 1pt off 4pt on 6pt off 4pt,inner sep=4mm, rectangle, rounded corners}};
 
 \tikzset{red dotted/.style={draw=red, line width=1pt,dash pattern=on 1pt off 4pt on 6pt off 4pt,inner sep=4mm, rectangle, rounded corners}};

 \node (first dotted box) [blue dotted, fit = (n1)  (n2) (n4)] {};
 \node (second dotted box) [red dotted,fit = (n2) (n3) (n5)] {};
\node at (first dotted box.north) [above, inner sep=3mm,color=blue] {\textbf{Phase-II: Post-Announcement}};
\node at (second dotted box.south) [below, inner sep=3mm, color=red] {\textbf{Phase I: Pre-Announcement}};
\draw[->, thick] (n3) -- (n2);
\draw[->, thick] (n3) -- (n5);
\draw[->, thick] (n2) -- (n1);
\draw[->, thick] (n2) -- (n4);
\end{tikzpicture}

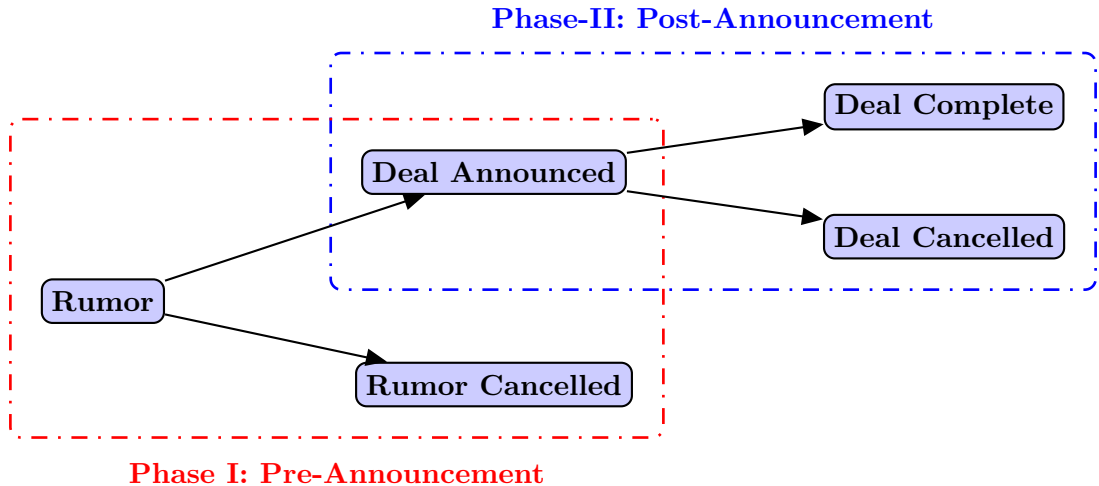
\captionof{figure}{M\&A Model Framework}
\label{plot:ma_framework}
\end{center}

This paper is organized as follows: In Section \ref{LR}, we go through the current literature on takeover success prediction. In Section \ref{Data}, we describe our data set and implement data analytic techniques to solve the problems with missing values, highly imbalanced classes and imbalanced features. In Section \ref{Methodology} and Section \ref{Results}, we introduce the overall methodology for both phases and compare our methodology with baseline models in this field. In Section \ref{Conclusion}, we summarize our study and explain our future extension plans. 
\section{Literature Review}\label{LR}
Merger and acquisitions have been one of the major research interests in finance. There are many studies that focus on different aspects of merger and acquisition process. In this section, we will first focus on the studies that prove run-ups before deal announcement. Then, we will go through the literature on takeover success prediction in terms of the important features and methodologies.

\subsection{Run-ups before announcement}

Numerous studies have attempted to identify the characteristics and the amount of the price run-ups prior to the announcement of the merger deals. Keown and Pinkerton \cite{keown1981merger} provided evidence of excess returns earned by investors in target firm stocks prior to the announcement. They examined the impact of trading on the target firms in advance of the announcement by focusing on the daily stock price movements of 194 successfully acquired firms. According to their observations, the share price of the target company starts increasing one-month before the announcement. In their study, they justified the pre-announcement run-ups as insider trading. While Jensen and Ruback \cite{jensen1983market} empirically observed that almost half of the abnormal returns associated with the deal announcements occur prior to the announcement,  the experiment results in Jarrell and Paulsen \cite{jarrell1989stock} indicate that about 40 percent of the takeover premium is in the form of pre-announcement run-up. Jarrell and Paulsen \cite{jarrell1989stock} also found that the presence of rumors is the strongest variable explaining pre-announcement run-ups. Pound and Zeckhauser \cite{pound1990clearly} examined the effects of rumors on the share price of the target companies. Their study shows that there has been approximately 7\% run-up over the 20 trading days before rumor publication. According to their study, the rumors are totally reflected in the market prices so that there will be no crucial impact on the share price of the target company on the date of the rumor publication. 18 out of 42 firms in their study announced a deal within a year of the rumor publication, but only two of them had the announcement within 50 days of the rumor publication. Draper and Paudyal \cite{draper1999corporate} observed that statistically significant excess returns start around three weeks prior to the announcement, and the cumulative excess returns reach 14.35 percent by the day of the announcement. In another study, Andrade et al. \cite{andrade2001new} showed that the share price of target firms increased by 23.8 percent in the window beginning 20 days prior to announcement and the day of the announcement. Gao and Oler \cite{gao2012rumors} observed that the stock prices of rumored firms drift down to their pre-rumor level over a 70-trading-day period after the initial price jump when a rumor is published, and only 12.6\% of rumored takeovers materialize into actual announcements within 70 trading days. Tang and Xu \cite{tang2016causes} investigated the increase in the share price of the target companies prior to deal announcements between 1981 and 2011. They observed that about one-third of the total price run-up occurs before announcements, and they found that these run-ups are significantly associated with insider trading instead of market anticipation or toehold acquisition. 

\subsection{Important features for takeover success prediction}

A number of empirical studies in the literature have attempted to identify the important features to be associated with takeover success prediction. In this section, we will go through these papers to explore the relationship between these features and the outcome of the takeover attempt. 

The previous literature has analyzed three types of information while predicting the takeover outcomes that are firm/deal information, market price information and risk arbitrageur information. Firm/deal information contains the information related to the firm and deal characteristics such as bid premium, target's management ownership, leverage, size and merger experience of the acquirer. The second type of information includes post-announcement price movement and trading volume of the target firm's shares. The last one explores the effect of arbitrageur's purchasing behavior on the prediction.

Hoffmeister and Dyl \cite{hoffmeister1981predicting} used takeover firm's financial condition at year-end prior to the announcement of the takeover attempts, target firm's vulnerability to takeover attempts, target firm's standing in its industry, and management attitude in tender offer in the prediction model. Their research indicated that management attitude and firm size are the most important variables among all variables they considered. Walkling \cite{walkling1985predicting} investigated the hypothesis that while bid premium and the payment of solicitation fees have positive impact on the probability of success, target management opposition and competing bids negatively affect the probability. Adelaja \cite{adelaja1999predicting} found that the target attitude has the greatest impact on the prediction model. Cudd and Duggal \cite{cudd2000industry} considered only the financial ratios as predictors. Baker and Savaşoglu \cite{baker2002limited} observed that hostility of the acquirer explains 8 percent of the variation in the outcomes and it is significant at 1\% level. According to Officer \cite{officer2007performance}, the existence of termination fees could improve the probability of merger completion as it would make withdrawal from the merger deal costly. Branch and Yang \cite{branch2003predicting} tested how payment methods and merger types influence the success of takeovers and later they showed that the deals with cash offers are more likely to succeed than those with stock swaps. Management attitude and the relative size of the target (target asset size/acquirer asset size) are found to be the most significant factors based on their stepwise model. In their later study \cite{branch2008note}, they discovered that arbitrage spread, management attitude(1 for friendly, 0 for hostile), transaction size and deal structure(1 for stock swaps, 0 for cash offers) are the most important predictors in their model. In consistence with other studies, Noro et al. \cite{noro2010takeover} also showed that management attitude is of great significance for prediction model for European market takeover attempts. In their  statistical analysis, Jetley and Ji \cite{jetley2010shrinking} identified payment structure, bid premium, management attitude, trading volume in the target company's stock one day after announcement and target company's market capitalisation as significant at 5\% level. Chakrabarti and Mitchell argued that the takeover success can be partially explained by their spatial characteristics. In order to incorporate graphical distance into their model, they considered the log of the miles between coordinates for the zip codes of the acquirer and target firm headquarters. Lee et al. \cite{lee2020unbalanced} investigated the impact of the financial and deal related predictors on the prediction of takeover outcomes. In contrast to other studies in the literature, they took the absolute difference between financial measures of the two companies and included them in their model as predictors. Their feature importance model indicated that termination fee clause dummy, existence of competing bids, management attitude and total asset difference have the greatest power on the predictions. 

To summarize, existing studies in the takeover success prediction consider different firm/deal information, market price information and risk arbitrageur information as predictors. Most of these studies conclude that management attitude, relative size of the target, termination fee, payment methods, bid premium and competing bids provide significant information while constructing prediction framework. Based on these studies, the resistance from the management of the target company, existence of termination fee and existence of competing bids would make the deal completion harder. As the relative size of the target company increases, the change that the target company is acquired decreases. Compared to the deals with stock swaps, deals with cash offers are more successful in terms of the completion.

\subsection{Literature on takeover success prediction methodologies}

There are a number of studies in the literature that have constructed takeover success prediction models. Most of the existing studies focus on implementing the logit and probit models with various features. Recently, there are also papers implementing neural networks and different machine learning algorithms. In this section, we will go through the papers which address the takeover success prediction problem with different methodologies. Table \ref{table:literature} provides an overview of the related papers and their prediction accuracy rates for overall, success and failure cases. 

The earlier research on this field benefited from either ordinary least squares or multivariate discriminant analysis to build a prediction framework for takeover outcomes. Hoffmeister and Dyl \cite{hoffmeister1981predicting} built multivariate discriminant analysis model using 17 independent variables including features on target firm's financial condition at year-end, target firm's vulnerability to takeover attempts, target firm's standing in its industry and management's reaction to takeover attempts. While the overall accuracy of the model was around 85\%, failure case accuracy was 28.6\%. Although several studies used these models for takeover success prediction, there are certain drawbacks using them for the prediction. First, the predicted probabilities from linear regression models can violate zero-one boundaries. Furthermore, building linear regression models with binary outcomes can lead to heteroskedasticity. 

In order to overcome the problems caused by linear regression models, logit and probit models became more prominent later in the literature on the takeover success prediction. The predicted probabilities from these models obey zero-one boundaries but still the decision boundary based on these models is a linear function of the predictor variables. Walkling \cite{walkling1985predicting}, Duggal and Millar \cite{duggal1994institutional}, Adelaja \cite{adelaja1999predicting}, Cudd and Duggal \cite{cudd2000industry}, Betton and Eckbo \cite{betton2000toeholds}, Branch and Yang \cite{branch2003predicting}, Bange and Mazzeo \cite{bange2004board}, Bhagat et al. \cite{bhagat2005tender}, Heron and Lie \cite{heron2006use}, Noro et al. \cite{noro2010takeover}, Jeon and Ligon \cite{jeon2011much}, Khrishnan and Masulis \cite{krishnan2013law}, Renneboog and Zhao \cite{renneboog2014director} and Chakrabarti and Mitchell \cite{chakrabarti2016role} built their prediction frameworks using the logit model. Adelaja \cite{adelaja1999predicting} constructed the model to predict the takeover success in food industry between 1985 and 1994 and Cudd and Duggal \cite{cudd2000industry} used financial ratios to predict the target outcomes in mining and manufacturing sectors. Branch and Yang \cite{branch2003predicting} utilized stepwise logistic regression to obtain a model with the most important features and assessed the fit of the model using Hosmer-Lemeshow test. Although most of the studies investigated the takeover attempts in US market, Noro et al. \cite{noro2010takeover} worked on identifying the determinants of takeover success on European market. Furthermore, Baker and Savaşoglu \cite{baker2002limited}, Officer \cite{officer2007performance} and Bodt et al. \cite{de2018empirical} employed probit model for takeover success prediction. Lin et al. \cite{lin2013option} proposed a completely different framework using option-cased approach. They defined the probability of takeover success as in-the-money probability. While selecting bid premium, relative size of the target, shares controlled by the bidder and payment methods as predictors in their logit and probit models, they only used stock prices and stock exchange ratios in order to predict takeover outcomes.

Although logit and probit models solve the problems arising from linear regression models, these models continue to use linear decision boundaries between the outcome and the predictors. It could limit the advantages of logit and probit models as some features may have nonlinear relationship with takeover outcomes that require nonlinear decision boundaries. Therefore, some recent studies have constructed their prediction models using neural networks and machine learning techniques. Branch et al. \cite{branch2008note} compared the performance of the neural networks with the traditional logit model. While constructing the neural network, they used a single hidden layer with 7 neurons and sigmoid activation function. Their empirical results indicate that neural networks outperforms the logistic regression in predicting failed takeover attempts. Zhang et al. \cite{zhang2012predicting} extended the work from \cite{branch2008note} by comparing the performance of the logit model with different machine learning algorithms consisting of neural networks, SVM with different kernels, decision tress, random forest and Adaboost. They used exactly the same data set in \cite{branch2008note}. According to their results, SVM with linear kernel and Adaboost with stump weak classifiers perform the best. In order to address the bias in the dominant class that is observed in most of the data sets, Lee et al. \cite{lee2020unbalanced} proposed a generalized logit framework and a context-specific cost-sensitive function. For this purpose, they used neural networks with 3 neurons in the single hidden layer and sigmoid activation function. The final model results are compared with logit, probit, weighted logit models and plain neural networks without cost-sensitive function. As an evaluation metric, overall accuracy and failure accuracy are chosen since misclassifying failed deals as successful deals can lead to much severe consequences than misclassfying successful deals as failed deals. Furenmo \cite{furenmo2020predicting} compared the performance of random forest with the performance of the logit model, and found that random forest model outperforms logistic regression.

\begin{table}[htbp]
\vskip\baselineskip 
\begin{center}
\begin{adjustwidth}{-0.5cm}{0cm}
\scalebox{0.70}{
\begin{threeparttable}
\begin{tabular}{c c|| c c c c| l|c c c| c c c}\hline
\textbf{Paper}& \textbf{Year} & \multicolumn{4}{c|}{\textbf{Data}}& \textbf{Methodology} &\multicolumn{3}{c|}{ \textbf{In-Sample}}&  \multicolumn{3}{c}{ \textbf{Out-of-Sample}}\\
&&{\small \textbf{Period}}&{\small \textbf{S}}&{\small \textbf{F}}&{\small \textbf{F/O}}&&{\small \textbf{O}}&{\small \textbf{S}}&{\small \textbf{F}}&{\small \textbf{O}}&{\small \textbf{S}}&{\small \textbf{F}}\\\hline
\cite{lee2020unbalanced} & 2020& 2009-2014 & 411 & 56 & 12\% & \textbf{neural network with cost}&\textbf{95.0\%}& -& \textbf{96.0\%}& \textbf{89.0\%}& -& \textbf{75.0\%} \\ 
 & &  &  &  &  & neural network&97.0\%& -& 77.0\%& 89.0\%& -& 38.0\% \\
 & &  &  &  &  & probit&93.0\%& -& 58.0\%& 89.0\%& -& 25.0\% \\
 & &  &  &  &  & weighted logit&88.0\%& -& 42.0\%& 88.0\%& -& 38.0\% \\
 & &  &  &  &  & logit&93.0\%& -& 63.0\%& 88.0\%& -& 38.0\% \\ 
\cite{furenmo2020predicting} & 2020 & 2000-2018 & 6070 & 1333 & 18\% &\textbf{random forest}& - & - & - & \textbf{85.2\%}& \textbf{98.3\%}& \textbf{28.7\%} \\
 &  &  &  &  & &LASSO logit& - & - & - & 84.2\%& 97.7\%& 26.2\% \\
  &  &  &  &  & &logit& - & - & - & 84.0\%& 97.2\%& 26.3\% \\
\cite{lin2013option} & 2013 & 2000-2007 & 66 & 38 & 36\% &\textbf{option-based approach}& \textbf{71.2\%}& \textbf{95.4\%}& \textbf{29.0\%}& -& -& - \\ 
 &  &  &  &  &  &probit& 54.8\%& 57.6\%& 50.0\%& -& -& - \\
  &  &  &  &  &  &logit& 54.8\%& 57.6\%& 50.0\%& -& -& - \\
\cite{zhang2012predicting} & 2012 & 1991-2004 & 1050 & 146 & 12\% &\textbf{SVM-linear kernel}& -& -& -& \textbf{91.6\%}& \textbf{97.1\%}& \textbf{54.7\%} \\ 
 &  &  &  & &  &neural network& -& -& -& 90.9\%& 96.7\%& 51.8\% \\ 
&  &  &  & &  &AdaBoost& -& -& -& 91.3\%& 97.4\%& 50.4\% \\ 
&  &  &  & &  &random forest& -& -& -& 90.9\%& 97.4\%& 47.4\% \\ 
&  &  &  & &  &logit& -& -& -& 91.0\%& 97.1\%& 50.3\% \\ 
\cite{noro2010takeover} & 2010& 1999-2008 & 664 & 68 & 12\% &logit & 89.3\%& 98.3\%& 23.0\%& -& -& - \\ 
\cite{branch2008note} &2008 & 1991-2004 & 1050 & 146 & 12\% & \textbf{neural network}& -& -& -& -& \textbf{98.2\%}& \textbf{58.0\%} \\
 & &  &  &  &  & logit& -& -& -& -& 97.2\%& 47.9\% \\
\cite{branch2003predicting} & 2003 & 1991-2001 & 86 & 13 & 13\% &logit& -& -& -& 88.9\%& 96.5\%& 38.5\% \\
\cite{cudd2000industry} & 2000 & 1987-1991 & 13 & 460 & 97\% &logit& -& -& -& 76.1\%& 7.7\%& 78.0\% \\
\cite{adelaja1999predicting} & 1999 & 1985-1994 & 30 & 5 & 14\% &logit& 85.7\%& 93.3\%& 40.0\%& -& -& - \\ 
\cite{walkling1985predicting} & 1985 & 1972-1977 & 110 & 48 & 30\% &logit& 79.6\%& 85.5\%& 63.9\%& 60.0\%& 55.3\%& 75.0\% \\
\cite{hoffmeister1981predicting} & 1981 & 1976-1979 & 88 & 29 & 25\% &MDA& 78.6\%& 83.9\%& 63.6\%& 84.8\%& 100.0\%& 28.6\% \\
\hline
\end{tabular} 
\begin{tablenotes}[para,flushleft]
When there are multiple methodologies used in a paper, we bold the results of the methodology that generates the best results. Note that S, F, and O are the abbreviations used for success, failure and overall, respectively. Finally, F/O indicates the proportion of the failure cases within the data.
\end{tablenotes}
\end{threeparttable}}
\end{adjustwidth}
\end{center}
\caption{M\&A Success/Failure Prediction Literature}
\label{table:literature}
\end{table}

\section{Data}\label{Data}
\begin{figure}[t]
   \begin{minipage}{0.5\textwidth}
     \centering
     \includegraphics[scale = 0.35]{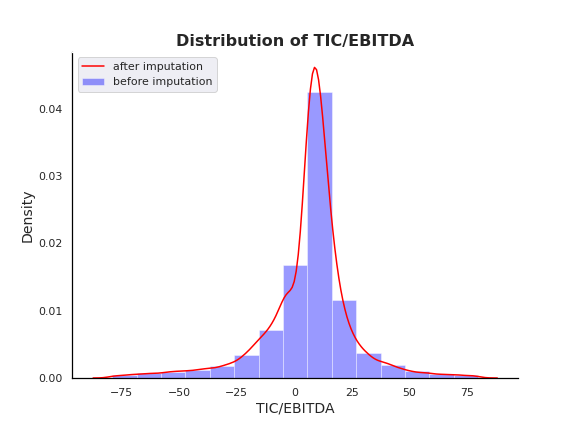}
   \end{minipage}\hfill
   \begin{minipage}{0.5\textwidth}
     \centering
     \includegraphics[scale = 0.35]{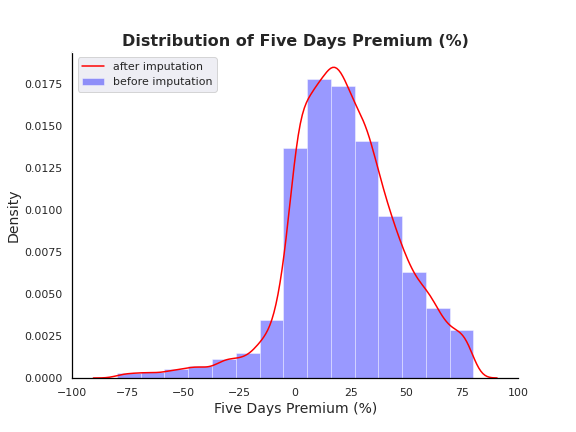}
   \end{minipage}
   \caption{Distribution of two features before/after imputation}\label{Fig:imputation}
\end{figure}

There are a number of data vendors providing deal information. In this study, we utilize the data from FactSet as it consists of financial data related to target and acquirer firms and deal characteristics such as method of payment, attitude, and so on. Moreover, FactSet also specifies when there is a rumor related to a deal. Our dataset consists of 19,006 unique M\&A deals with transaction values greater than \$1 million and public target ownership for the announcement time horizon from January 1, 2001 to October 30, 2020. After removing the variables that do not provide any information for at least $30\%$ of the deals, we end up with 52 numerical, 40 binary, and 11 categorical variables. Our numerical variables consists of financial ratios of target companies such as TIC/EBITDA, and the share prices of the target companies before and after deal announcement with different periods like a year prior or a week later share prices. Categorical and binary variables mostly consists of deal characteristics such as the region of the target company, the industry of the acquirer company, management attitude, method of payment, and existence of competing bidders. As there are still variables with missing values, we further implement \textit{K Nearest Neighbour} (\texttt{kNN}) algorithm for data imputation. This algorithm finds the k closest data points to a given data point using a distance metric and then labels this point based on the majority of the labels among these k closest points. After experimenting with different k values, we chose k equals to five. Figure \ref{Fig:imputation} shows the distribution of two selected features (TIC/EBITDA and Five Days Premium (\%)) from the dataset before and after applying \texttt{kNN} algorithm. As it is seen, the distributions of the variables stay almost the same after the data imputation. In order to use categorical features in modeling, we need to represent them numerically. Therefore, we apply one-hot encoding technique for converting categorical variables into multiple binary variables. Basically, each category is expressed as a binary variable. After this step, our data set contains 52 numerical and 108 binary variables. 

\FloatBarrier

We also aim at analyzing the impact of sentiment scores of the target companies on the accuracy of our model framework. With this regard, we benefit from the sentiment scores provided by Refinitiv MarketPsych. These scores range between -1 and 1, and a larger absolute value indicates a stronger view about the company. In order to explore the effects of the deal announcement, we use the sentiment scores for the time period from 90 days prior to announcement to 31 days after the announcement. 

Once we merge the data sets from FactSet and Refinitiv MarketPsych, our final dataset includes 17,440 unique deals. We further divide this dataset into two: training data and testing data. Training data contains 16,525 deals prior to 2019, and $80.84\%$ of these deals are completed. On the other hand, the testing data contains 915 deals that are announced starting 2019, and $79.89\%$ of them are completed. 

\begin{figure}[t]
\captionsetup{font=scriptsize,labelfont=scriptsize}
   \begin{minipage}{0.5\textwidth}
     \centering
     \includegraphics[scale = 0.35]{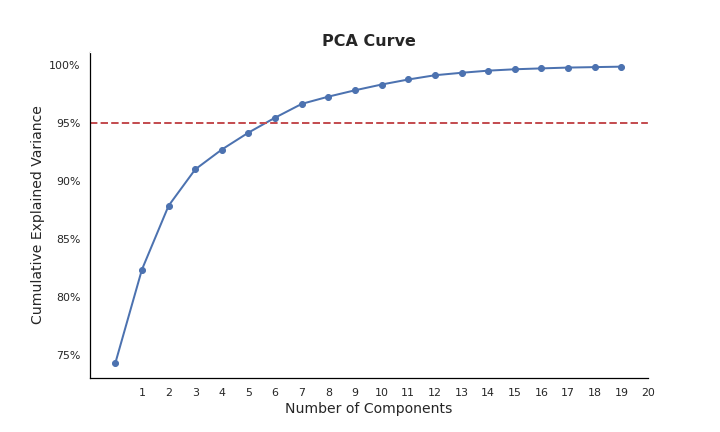}
     \caption{PCA on numerical features}\label{Fig:pca}
   \end{minipage}\hfill
   \begin{minipage}{0.5\textwidth}
     \centering
     \includegraphics[scale = 0.35]{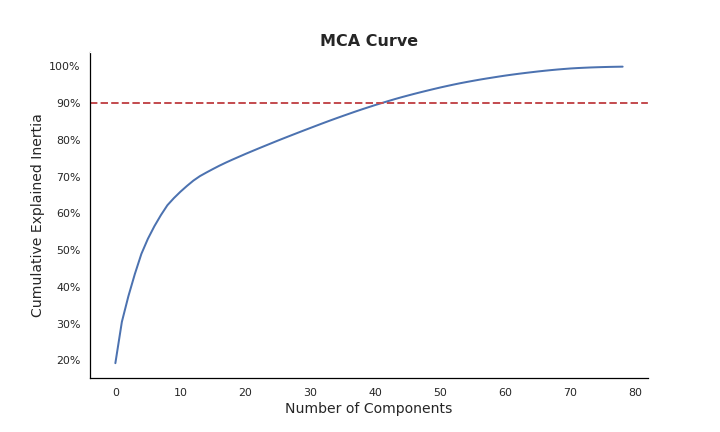}
     \caption{MCA on categorical features}\label{Fig:mca}
   \end{minipage}
\end{figure}

When the data preprocessing is done, we obtain 52 numerical variables, 108 encoded categorical variables, and 121 sequential sentiment scores. However, high-dimensional data increases the computational workload and may result in overfitting in the applied algorithm. Therefore, a variety of dimensionality reduction techniques are used based on the type of the data. For numerical features, we employ \textit{Principal Component Analysis} (\texttt{PCA}) (for further information \cite{pearson1901liii}, \cite{shlens2014tutorial}) that is an orthogonal linear transformation of the data to another coordinate system. In this method, we start with the vectors of all available numerical features and project them into a lower dimensional subspace by maximizing the variance. The first direction (principal component) will be the direction along which the projections have the highest variance. Similarly, the second principal component is the orthogonal direction to the first principal component along which the variance is maximized, and other principal components are extracted in the same manner. Mathematically speaking, we assume that $\mathbf{X}$ is $m\times n$ matrix where $m$ represents the number of numerical features and $n$ is the number of data points. We want to obtain a mapping $\mathbf{PX = Y}$ such that $\mathbf{P}$ is orthonormal and the covariance matrix $\mathbf{C=\frac{1}{n-1}YY^T}$ is diagonalized. Therefore, 
\begin{equation*}
    \mathbf{C=\frac{1}{n-1}YY^T=\frac{1}{n-1}P(XX^T)P^T=\frac{1}{n-1}P(P^T\Lambda P)P^T=\frac{1}{n-1}\Lambda}
\end{equation*}
In equation, $\mathbf{XX^T}$ is replaced by $\mathbf{P^T\Lambda P}$ because eigenvector decomposition of symmetric matrices can be written as $\mathbf{Q\Lambda Q^T}$ where $\mathbf{Q}$ is matrix of orthogonal eigenvectors, and $\mathbf{\Lambda}$ is diagonal matrix with corresponding eigenvalues. For this purpose, we take $\mathbf{Q=P^T}$, and the rest of the equation follows as stated. Therefore, the rows of $\mathbf{P}$ will constitute the principal components of $\mathbf{X}$ and the entries of $\mathbf{\Lambda}$ will be the explained variance of the $\mathbf{X}$ along the corresponding direction. 
Figure \ref{Fig:pca} illustrates the change in the cumulative explained variance as the number of principal components increases. As it is seen, we can use the first 20 principal components to represent our 52 dimensional numerical variables without losing too much information. 

\FloatBarrier

A similar methodology to \texttt{PCA}, which is called \textit{Multiple Correspondence Analysis} (\texttt{MCA}) (for further information \cite{greenacre1984theory},\cite{greenacre2006multiple}), is applied to categorical variables for dimensionality reduction. In \texttt{MCA}, loss of information is described by inertia that is proportional to $\chi^2$ statistic, and it measures the dispersion of the data around the expected values. We assume that we have \textbf{I} unique deals and \textbf{Q} categorical variables; each variable has $J_Q$ levels. Then, \textbf{X} will be $I\times J$ indicator matrix (one hot encoding) where $J=\sum_Q J_Q$ and $\mathbf{P=N^{-1}X}$ will be the correspondence matrix where N stands for the grant total of the table. For marginal relative frequencies, we calculate \textbf{r} (vector of row totals of $\mathbf{P}$) and \textbf{c} (vector of column totals of $\mathbf{P}$) and their corresponding diagonal matrices $\mathbf{D_r}$ and $\mathbf{D_c}$, respectively. Then, we centralize the data by calculating $\mathbf{P-rc^T}$ and standardize the centralized data as $\mathbf{Y=D_r^{-1/2}(P-rc^T) D_c^{-1/2}}$. Using Singular Value Decomposition, $\mathbf{Y}$ can be written as $\mathbf{U\Sigma V^T}$ where $\mathbf{\Sigma}$ is the diagonal matrix with singular values in descending order, and $\mathbf{\Sigma^2}$ is the diagonal matrix of eigenvalues of $\mathbf{SS^T}$. These eigenvalues are defined as \textit{principal inertia} in the context of correspondence analysis, and their sum will constitute the \textit{total inertia}. Furthermore, principal components of categorical variables are calculated as $\mathbf{Z=D_c^{-1/2}V\Sigma}$. For a n-dimensional map, we will consider the first n columns of $\mathbf{Z}$. The proportion of explained inertia by n-dimension will be the sum of eigenvalues corresponding to these columns over the total inertia. Figure \ref{Fig:mca} shows the pattern of the cumulative explained inertia with the increase in the number of principal components. We can explain more than $90\%$ of the inertia with 45 components, which means we can illustrate our 108 encoded categorical variables with 45 components without losing too much information.

\begin{figure}[t]
\centering
\begin{adjustbox}{max totalsize={.8\textwidth}{.7\textheight},center}
\begin{tikzpicture}[auto, thick, >=triangle 45]]
\matrix [column sep={5mm}, row sep=5mm] {
    \node(n1) [draw, shape=rectangle,fill=red!40, minimum height=70mm,minimum width=2em]{}; &
    \node(n2)  [draw, trapezium,rotate=270,fill=blue!40, minimum height=3.5em,minimum width=1.5em,trapezium angle=40,inner xsep=5pt]{}; & 
    \node(n3) [draw, shape=rectangle,fill=yellow!40, minimum height=20mm,minimum width=2em] {}; &
    \node(n4)  [draw, trapezium,rotate=90,fill=blue!40, minimum height=3.5em,minimum width=2em,trapezium angle=40,inner xsep=5pt]{}; & 
    \node(n5) [draw, shape=rectangle,fill=red!40, minimum height=70mm,minimum width=2em]{};  \\
    \node(n6) {\textbf{\begin{tabular}{c} Input \\ Sequence \end{tabular}}}; &
    \node(n7) {\textbf{Encoder}}; &
    \node(n8) {\textbf{\begin{tabular}{c} Lower \\ Dimensional \\Embedding \end{tabular}}}; &
    \node(n9) {\textbf{Decoder}}; &
    \node(n10) {\textbf{\begin{tabular}{c} Reconstructed \\ Sequence \end{tabular}}}; \\
};

 \tikzset{blue dotted/.style={draw=blue, line width=1pt,dash pattern=on 1pt off 4pt on 6pt off 4pt,inner sep=2mm, rectangle, rounded corners}};
 
 \node (first dotted box) [blue dotted, fit = (n1)  (n6) (n3) (n8), xshift=-0.4cm] {};

\node at (first dotted box.north) [above, inner sep=3mm,color=blue] {\textbf{Dimensionality Reduction}};

\draw[->, thick] (n1) -- (n2);
\draw[->, thick] (n2) -- (n3);
\draw[->, thick] (n3) -- (n4);
\draw[->, thick] (n4) -- (n5);
\end{tikzpicture}
\end{adjustbox}
\captionof{figure}{LSTM Autoencoder}
\label{plot:lstm_autoencoder}
\end{figure}
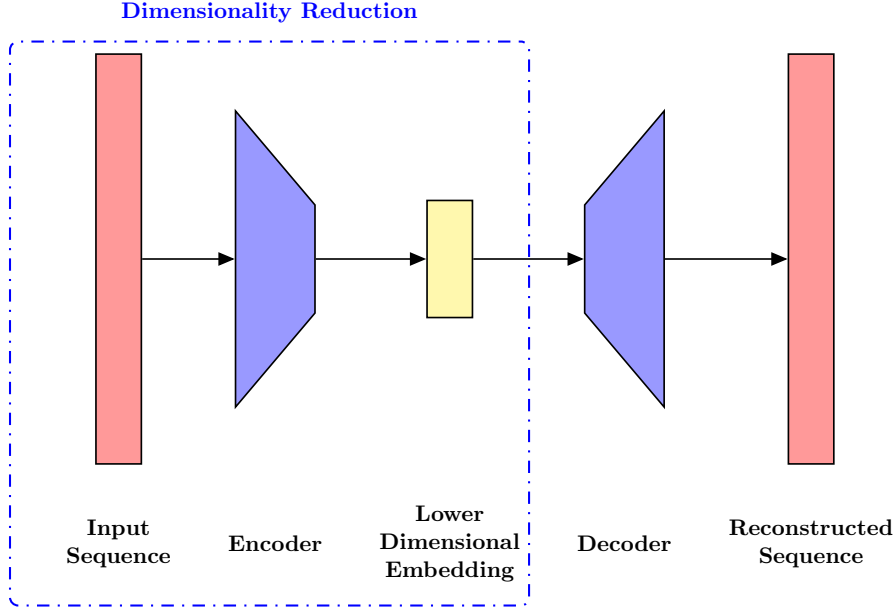

Furthermore, we have 121 days sequential sentiment scores for the target company of each deal. In order to use this temporal dataset together with non-temporal variables, we propose \textit{LSTM Autoencoders} (for further information \cite{srivastava2015unsupervised}) that is a type of autoencoder where encoder and decoder are both \texttt{LSTM} (long short-term memory) models. Figure \ref{plot:lstm_autoencoder} provides a framework of \texttt{LSTM} autoencoders. We assume that $\mathbf{X=(X_1,\cdots,X_T)}$ is the input sequence where $\mathbf{T}$ is the length of the sequence. Accordingly, encoder \texttt{LSTM} takes the input sequence $\mathbf{X}$ and reduce it into a lower dimensional vector $\mathbf{Z\in \mathbb{R}^K}$ where $\mathbf{K<T}$. Then, decoder \texttt{LSTM} takes $\mathbf{Z}$ as input and reconstruct the input sequence as $\mathbf{\hat{X}}$. As the purpose of the overall model is to reconstruct the input sequence, mean Euclidean distance between $\mathbf{X}$ and $\mathbf{\hat{X}}$ is used as the loss function while training the model. By employing this framework, we can compress our high-dimensional sequential sentiment scores without losing too much information and then use the lower dimensional embedding together with non-temporal variables in different models such as supervised feedforward neural networks. 

As mentioned earlier, our dataset is highly imbalanced because around 80\% of the deals are completed and 20\% of the deals are cancelled. In addition, misclassifying a cancelled deal as a completed deal may lead to high costs for investors. Therefore, it is crucial to predict the cancelled deals as accurate as possible. To address the class imbalance, we employ \textit{Synthetic Minority Oversampling Technique} (\texttt{SMOTE}) (for further information \cite{chawla2002smote}). This algorithm generates synthetic examples by operating in the feature space. Assume that $\mathbf{S_{min}}$ is the sample set of feature vectors of minority class. Then, the k-nearest neighbors of each feature vector $\mathbf{x\in S_{min}}$ are obtained by calculating the Euclidean distance between each of them. Depending on the sampling rate chosen, a number of nearest neighbors are selected randomly. If $\mathbf{y}$ is one of such neighbors, a new synthetic example will be generated as $\mathbf{x+(y-x)u}$, where $\mathbf{u}$ is a standard uniform random variable. It indicates that synthetic examples are generated along the line segment joining the feature vectors with their randomly selected nearest neighbors. In our problem, we employ this algorithm depending on the experiment setting.

\section{Methodology}\label{Methodology}
In this section, we first introduce different evaluation metrics and loss functions that we experiment with in Section \ref{Results}. We then propose three different classification frameworks for takeover success prediction. 

\subsection{Evaluation Metrics}
There are a number of evaluation metrics for binary classification problems. When the dataset is balanced, we usually use the overall accuracy of the predictions as the performance metric.  Equation \ref{accuracy} provides the mathematical formula of the overall accuracy. In this equation, \texttt{TP}, \texttt{TN}, \texttt{FP}, and \texttt{FN} represent the number of positive cases correctly classified as positive, the number of negative cases correctly classified as negative, the number of negative cases incorrectly classified as positive, and the number of positive cases incorrectly classified as negative, respectively.  
\begin{equation}
    \text{Accuracy} = \frac{TP+TN}{TP+FP+TN+TP} \label{accuracy}
\end{equation}
The problem arises when the dataset is highly imbalanced since the accuracy will be high even if we label each sample with the label of the majority class. In that case, we determine the evaluation criteria depending on our priority in the model. The evaluation metrics we consider in this study are following:
\begin{align}
   & \text{Precision} = \frac{TP}{TP+FP} \label{precision}  \\
  &  \text{Recall} =\frac{TP}{TP+FN} \label{recall} \\
  &  \text{F1-score} = \frac{2\times \text{Precision} \times \text{Recall}}{\text{Precision}+\text{Recall}} \label{f-score}
\end{align}
To start with, our dataset is highly imbalanced as around 20\% of the deals are cancelled. We label the completed deals as 0 and cancelled deals as 1. Equation \ref{precision} shows that \textit{precision} is the ratio of the number of correctly classified positive cases to the total number of positive predictions. It is a good metric when we care more about reducing the false positive predictions. Equation \ref{recall} shows that \textit{recall } is the ratio of the number of correctly classified positive cases to the total number of positive cases. Higher recall implies lower number of false negative predictions. In our problem, the cost of classifying a cancelled deal as a completed deal (\texttt{FN}) is much higher than the cost of classifying a completed deal as a cancelled deal (\texttt{FP}). Therefore, recall is more meaningful than precision as an evaluation metric. Furthermore, \textit{F1-score} is the harmonic mean of the recall and the precision as in Equation \ref{f-score}. It focuses on decreasing both false negatives and false positives. Therefore, it also provides a good insight while evaluating the performance of the models. 

We further utilize \textit{Receiver Operator Characteristic} (\texttt{ROC}) Curve and \textit{Precision-Recall} (\texttt{PR}) Curve to evaluate the model performance. Both curves are commonly used to exhibit the results of binary classification problems. \texttt{ROC} curve is a graphical illustration of the trade-off between true positive rate (recall) and false positive rate ($\frac{FP}{FP+TN}$) for different thresholds that separate two classes. \texttt{AUROC} is the area under \texttt{ROC} curve that provides a single score that indicates the benefit of using the model. However, it is shown in Davis and Goadrich \cite{davis2006relationship} that \texttt{ROC} curve may not be powerful when the dataset is highly imbalanced. Even if the model results in a large number of false positives, false positive rate will remain low due to the large number of negative cases. In turn, \texttt{ROC} curve will not be able to show the performance of the model. On the other hand, \texttt{PR} curve is used to show the trade-off between precision and recall for every possible cut-off point. As opposed to false positive rate, precision compares the true positives with false positives. Therefore, it provides more information regarding the performance of the different models. Similarly, \texttt{AUPR} is the area under \texttt{PR} curve that gives a single score regarding the performance of the model across different thresholds. 

\subsection{Loss Functions}

Cross-entropy is the most commonly used loss function for binary classification problems. Assume that $p$ is the true label and $q$ is the predicted probability value of the true label. The cross-entropy function $H(p,q)$ is used to measure the dissimilarity between $p$ and $q$. Its value increases as the predicted probability diverges from the true label. Equation \ref{cross_entropy} provides the mathematical expression of the loss function that takes the average of the binary cross-entropy loss function over the sample with size $N$.
\begin{align}
    \text{Cross-Entropy Loss} &= \frac{1}{N}\sum_{n=1}^{N} H(p_n,q_n) \nonumber \\
    &= -\frac{1}{N}\sum_{n=1}^{N} \left[p_n \log(q_n) + (1-p_n) \log(1-q_n)\right] \label{cross_entropy}
\end{align}

When the dataset is highly imbalanced, simply using binary cross-entropy function as a loss function may not be good enough. Therefore, we also utilize focal, F1, and Tversky loss functions to address class imbalance within the loss function. Focal loss function $FL(p,q)$ is introduced by Lin et al. \cite{lin2017focal}. By adding a modulating factor with a tunable focusing parameter $\gamma>0$, it dynamically scales the binary cross-entropy function such that the relative loss of well-classified examples ($(q>0.5$ for $p=1$ and $1-q>0.5$ for $p=0$) is decreased and more focus is put on misclassified examples. Equation \ref{focal_loss} gives the mathematical expression of the focal loss function where the sample size is $N$. 
\begin{align}
    \text{Focal Loss} &= \frac{1}{N}\sum_{n=1}^{N} FL(p_n,q_n) \nonumber \\
    &= -\frac{1}{N}\sum_{n=1}^{N} \left[p_n (1-q_n)^{\gamma}\log(q_n) + (1-p_n)q_n^{\gamma} \log(1-q_n)\right] \label{focal_loss}
\end{align}

F1 score is a good evaluation metric for binary classification problems. However, 1 - F1 score cannot be used as a loss function directly as it is non-differentiable. By utilizing the prediction probabilities instead of the predicted labels, we can obtain a differentiable loss function. Equation \ref{f1_loss} shows the mathematical representation of the F1 loss function.
\begin{align}
    \text{F1 Loss} &= 1 - \frac{\sum_{n=1}^{N} p_n q_n}{\sum_{n=1}^{N} \left[p_n q_n + 0.5 (1-p_n)(q_n) + 0.5 p_n(1-q_n)\right]} \label{f1_loss}
 \end{align}   
 
 Tversky loss function is introduced by Salehi et al. \cite{salehi2017tversky}. It adjusts the trade-off between false positives and false negatives by controlling the constants $\alpha$ and $\beta$. Equation \ref{tversky_loss} provides the mathematical representation of Tversky loss function. In this equation, $\sum_{n=1}^{N} p_n q_n$, $\sum_{n=1}^{N} (1-p_n)(q_n)$, and $\sum_{n=1}^{N} p_n(1-q_n)$ represents \texttt{TP}, \texttt{FP}, and \texttt{FN}, respectively. When $\alpha$ and $\beta$ are both equal to 0.5, Tversky loss functions simply becomes F1 loss function. 
\begin{align}
    \text{Tversky Loss} &= 1 - \frac{\sum_{n=1}^{N} p_n q_n}{\sum_{n=1}^{N} \left[p_n q_n + \alpha (1-p_n)(q_n) + \beta p_n(1-q_n)\right]} \label{tversky_loss}
 \end{align}

\subsection{Models without sentiment scores}
\begin{figure}[t]
\centering
\begin{adjustbox}{max totalsize={.55\textwidth}{.4\textheight},center}
\begin{tikzpicture}[auto, thick, >=triangle 45]
\matrix [column sep={0mm}, row sep=15mm] {
    \node(n1) [draw, shape=rectangle,rounded corners,fill=red!40, minimum height=3.5em,minimum width=35mm]{\textbf{\begin{tabular}{c} Numerical \\ Data \end{tabular}}}; &&
    \node(n2)  [draw, shape=rectangle,rounded corners,fill=blue!40, minimum height=3.5em,minimum width=35mm]{\textbf{\begin{tabular}{c} Categorical \\ Data \end{tabular}}};\\
    \node(n3) [draw, shape=rectangle,rounded corners,fill=red!20, minimum height=3.5em,minimum width=35mm] {\textbf{PCA}};&& \node(n4) [draw, shape=rectangle,rounded corners,fill=blue!20, minimum height=3.5em,minimum width=35mm] {\textbf{MCA}};\\
    &\node(n5) [draw, dashed,shape=rectangle,rounded corners, minimum height=3.5em,minimum width=35mm] {\textbf{SMOTE}};&\\
    & \node(n6) [draw, shape=rectangle,rounded corners,fill=green!30, minimum height=3.5em,minimum width=35mm] {\textbf{FFNN}};&\\
    & \node(n7) [] {\textbf{\begin{tabular}{c} Deal \\ Status \end{tabular}}};&\\
};

\draw[->, thick] (n1) -- (n3);
\draw[->, thick] (n2) -- (n4);
\draw[->, thick] (n3) -- (n5);
\draw[->, thick] (n4) -- (n5);
\draw[->, thick] (n5) -- (n6);
\draw[->, thick] (n6) -- (n7);
\end{tikzpicture}
\end{adjustbox}
\captionof{figure}{\nth{1} Classification Framework}
\label{plot:classification1}
\end{figure}
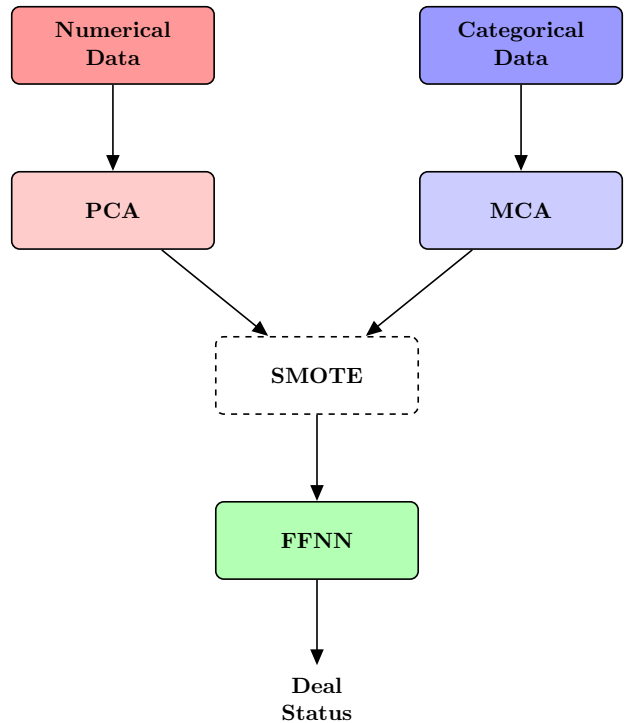

Figure \ref{plot:classification1} illustrates our initial classification framework. As detailed in Section \ref{Data}, our initial dataset consists of 52 numerical and 108 one-hot-encoded categorical variables. We employ \texttt{PCA} algorithm to represent the variables with 20 dimensions. Similarly, the dimension of the categorical variables is reduced to 45 dimensions by applying \texttt{MCA} algorithm. In the next step, \texttt{SMOTE} algorithm is implemented on the reduced dataset to balance the number of classes by oversampling the minority class. The reduced and oversampled dataset is then passed to a Feedforward Neural Network (\texttt{FFNN}) in order to predict the success/failure of the deals.

\FloatBarrier

\subsection{Models with sentiment scores}
\begin{figure}[t]
\centering
\begin{adjustbox}{max totalsize={.8\textwidth}{.8\textheight},center}
\begin{tikzpicture}[auto, thick, >=triangle 45]
\matrix [column sep={25mm}, row sep=15mm] {
    \node(n1) [draw, shape=rectangle,rounded corners,fill=red!40, minimum height=3.5em,minimum width=35mm]{\textbf{\begin{tabular}{c} Numerical \\ Data \end{tabular}}}; &
    \node(n2)  [draw, shape=rectangle,rounded corners,fill=blue!40, minimum height=3.5em,minimum width=35mm]{\textbf{\begin{tabular}{c} Categorical \\ Data \end{tabular}}}; & \node(n3) [draw, shape=rectangle,rounded corners,fill=yellow!40, minimum height=3.5em,minimum width=35mm] {\textbf{\begin{tabular}{c} Temporal \\ Data \end{tabular}}};\\
    \node(n4) [draw, shape=rectangle,rounded corners,fill=red!20, minimum height=3.5em,minimum width=35mm] {\textbf{PCA}};& \node(n5) [draw, shape=rectangle,rounded corners,fill=blue!20, minimum height=3.5em,minimum width=35mm] {\textbf{MCA}};& \node(n6) [draw, shape=rectangle,rounded corners,fill=yellow!20, minimum height=3.5em,minimum width=35mm] {\textbf{\begin{tabular}{c} LSTM \\ Autoencoder \end{tabular}}};\\
    &\node(n9) [draw, dashed,shape=rectangle,rounded corners, minimum height=3.5em,minimum width=35mm] {\textbf{SMOTE}};&\\
    & \node(n7) [draw, shape=rectangle,rounded corners,fill=green!30, minimum height=3.5em,minimum width=35mm] {\textbf{FFNN}};&\\
    & \node(n8) [] {\textbf{\begin{tabular}{c} Deal \\ Status \end{tabular}}};&\\
};

\draw[->, thick] (n1) -- (n4);
\draw[->, thick] (n2) -- (n5);
\draw[->, thick] (n3) -- (n6);
\draw[->, thick] (n4) -- (n9);
\draw[->, thick] (n5) -- (n9);
\draw[->, thick] (n6) -- (n9);
\draw[->, thick] (n9) -- (n7);
\draw[->, thick] (n7) -- (n8);
\end{tikzpicture}
\end{adjustbox}
\captionof{figure}{\nth{2} Classification Framework}
\label{plot:classification2}
\end{figure}
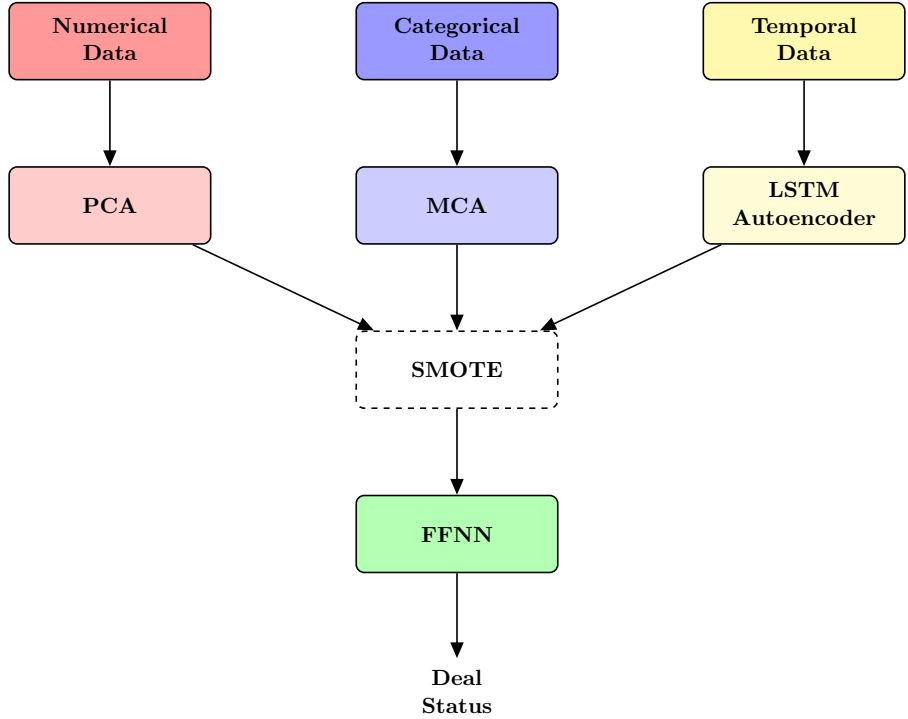

In addition to numerical and categorical variables representing financial information of the firms and deal characteristics, we further introduce sentiment scores of target companies into our model. Figure \ref{plot:classification2} illustrates the initial framework we propose with sequential sentiment scores. Similar to Figure \ref{plot:classification1}, we employ \texttt{PCA} on numerical variables and \texttt{MCA} on one-hot-encoded categorical variables to reduce the dimensionality of the dataset obtained from FactSet. At the same time, we implement \texttt{LSTM Autoencoder} on our temporal dataset to represent it with a lower dimensional embedding. In this model, we reduce 121-days sequential sentiment scores into a 5 dimensional embedding. Then, we merge all datasets and obtain a 70-dimensional final dataset. In order to address class imbalance, \texttt{SMOTE} algorithm is applied on the overall reduced dataset. A Feedforward Neural Network then takes the reduced and oversampled dataset as input and outputs the status of the deal.

\FloatBarrier

\begin{figure}[t]
\centering
\begin{adjustbox}{max totalsize={.8\textwidth}{.8\textheight},center}
\begin{tikzpicture}[auto, thick, >=triangle 45]
\matrix [column sep={1mm}, row sep=15mm] {
    \node(n1) [draw, shape=rectangle,rounded corners,fill=red!40, minimum height=3.5em,minimum width=35mm]{\textbf{\begin{tabular}{c} Numerical \\ Data \end{tabular}}}; &;&
    \node(n2)  [draw, shape=rectangle,rounded corners,fill=blue!40, minimum height=3.5em,minimum width=35mm]{\textbf{\begin{tabular}{c} Categorical \\ Data \end{tabular}}}; &;&;\\
    \node(n3) [draw, shape=rectangle,rounded corners,fill=red!20, minimum height=3.5em,minimum width=35mm] {\textbf{PCA}};&;& \node(n4) [draw, shape=rectangle,rounded corners,fill=blue!20, minimum height=3.5em,minimum width=35mm] {\textbf{MCA}};&;& \node(n5) [draw, shape=rectangle,rounded corners,fill=yellow!40, minimum height=3.5em,minimum width=35mm] {\textbf{\begin{tabular}{c} Temporal \\ Data \end{tabular}}};\\
    &\node(n6) [draw, dashed,shape=rectangle,rounded corners, minimum height=3.5em,minimum width=35mm] {\textbf{SMOTE}};&;&;&;\node(n7) [draw, dashed,shape=rectangle,rounded corners, minimum height=3.5em,minimum width=35mm] {\textbf{SMOTE}};\\
    & \node(n8) [draw, shape=rectangle,rounded corners,fill=green!30, minimum height=3.5em,minimum width=35mm] {\textbf{FFNN}};&;&;&;\node(n9) [draw, shape=rectangle,rounded corners,fill=cyan!30, minimum height=3.5em,minimum width=35mm] {\textbf{LSTM}};\\
    &;& \node(n10) [draw, shape=rectangle,rounded corners,fill=green!30,minimum width=35mm, minimum height=3.5em] {\textbf{FFNN}};&\\
    &;& \node(n11) [] {\textbf{\begin{tabular}{c} Deal \\ Status \end{tabular}}};&;\node(n12) [minimum width=35mm] {};\\
};

\draw[->, thick] (n1) -- (n3);
\draw[->, thick] (n2) -- (n4);
\draw[->, thick] (n3) -- (n6);
\draw[->, thick] (n4) -- (n6);
\draw[->, thick] (n5) -- (n7);
\draw[->, thick] (n6) -- (n8);
\draw[->, thick] (n7) -- (n9);
\draw[->, thick] (n8) -- (n10);
\draw[->, thick] (n9) -- (n10);
\draw[->, thick] (n10) -- (n11);
\end{tikzpicture}
\end{adjustbox}
\captionof{figure}{\nth{3} Classification Framework}
\label{plot:classification3}
\end{figure}
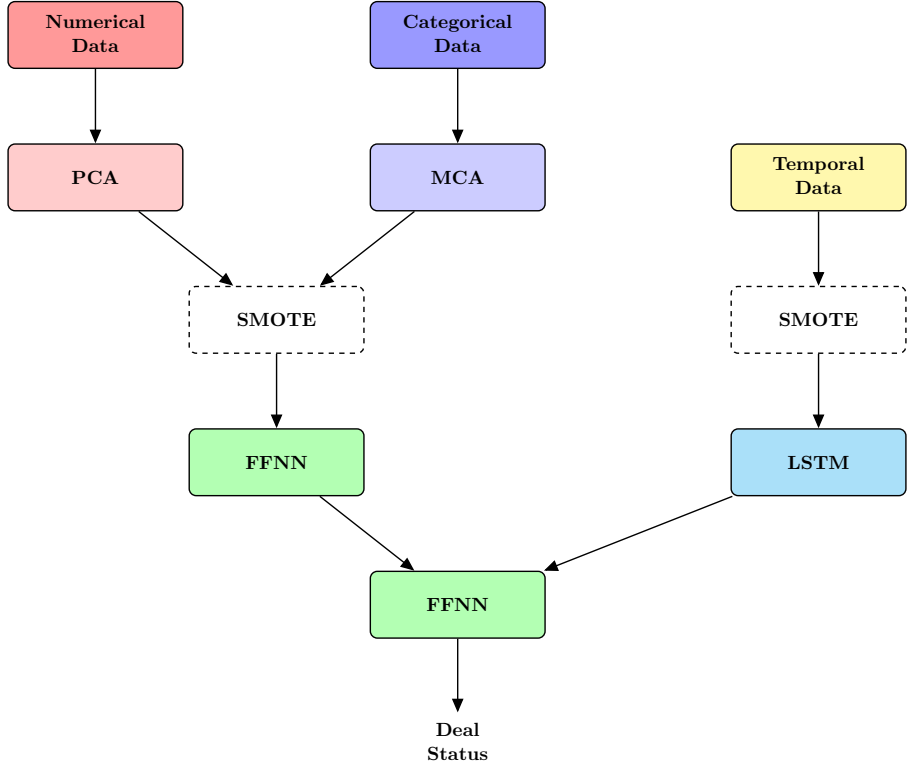

We further propose a second classification framework including sentiment scores. Figure \ref{plot:classification3} gives the illustration of the model. As in the previous models, we continue implementing \texttt{PCA} and \texttt{MCA} on numerical and categorical variables, respectively. Then, we handle class imbalance problem by applying \texttt{SMOTE} algorithm on the reduced dataset obtained from numerical and categorical variables and the original dataset of 121-days sentiment scores. The reduced and oversampled dataset of numerical and categorical variables is then passed to a Feedforward Neural Network, and the oversampled sentiment scores are passed to an \texttt{LSTM RNN}. \texttt{LSTM RNN} reads the sentiment scores one at a time from beginning to the end, and it outputs the final internal state. The output of the first \texttt{FFNN} and the output of the \texttt{LSTM} are then concatenated, and fed into the second \texttt{FFNN}, which will predict the status of the deal.

\FloatBarrier

\section{Results}\label{Results}
In previous section, we introduce three classification frameworks that we use in this study. For all models, there are a number of hyperparameters that needs to be tuned for the best results. For the tuning, we use Bayesian Optimization framework in order to decide on the optimal structure given a range for all hyperparamaters. 

Table \ref{table:result1} shows the results of the first classification framework, in which we do not include sentiment scores in the model framework. We compared our results against the logit and weighted logit models as they are the most common methods used for takeover success prediction problem. We categorized the results based on the evaluation metric and existence of oversampling algorithm. According to  the results of the Bayesian Optimization hyperparameter tuning, one hidden layer with 64 neurons is used for \texttt{NN-Recall} model, where we use \texttt{selu} (scaled exponential linear unit) activation function and binary cross-entropy loss function. \texttt{NN-Accuracy} model is defined by two hidden layers with 128 and 8 neurons, respectively. \texttt{Relu} (rectified linear unit) activation function is used in both layers, and binary cross-entropy function is chosen as the loss function. Similarly, two hidden layers with 256 and 8 neurons with \texttt{relu} activation function are used for \texttt{NN-F1} model. In this model, F1 Loss, which is defined in Equation \ref{f1_loss}, is chosen as the loss function. Our initial results show that we achieve higher accuracy levels compared to baseline models on the out-of-sample data without sacrificing recall too much, and we also gain a level of improvement in all other metrics. In order to improve model performance further, we then apply \texttt{SMOTE} algorithm and use Bayesian Optimization framework on the oversampled dataset. Accordingly, one hidden layer with 8 neurons and \texttt{elu} activation function is chosen for \texttt{SMOTE-NN-Recall} model. We use Focal Loss, which is defined in Equation \ref{focal_loss}, as the loss function. \texttt{SMOTE-NN-Accuracy} model is built with 1 layer with 256 neurons and \texttt{relu} activation function, and binary cross-entropy is used as the loss function. Moreover, one hidden layer with 8 neurons and \texttt{selu} activation function is used for \texttt{SMOTE-NN-F1} model, and F1 Loss is chosen as the loss function. As results indicate, \texttt{SMOTE-NN-Recall} model achieves highest recall value compared to other models and benchmark models in Table \ref{table:result1}. However, accuracy decreases in order to obtain the recall level. According to Table \ref{table:result1}, the best overall model is \texttt{SMOTE-NN-F1} as the trade-off between accuracy and recall is more balanced both for in-sample and out-of-sample results. The model is also pretty generalized as the results are consistent between in-sample and out-of-sample. 

\begin{table}[htbp]
\vskip\baselineskip 
\begin{center}
\begin{adjustwidth}{-1.6cm}{0cm}
\scalebox{0.68}{
\begin{tabular}{ l|| c c c c c c|c c c c c c}\hline
\textbf{Methodology} & \multicolumn{6}{c|}{\textbf{In-Sample}}&\multicolumn{6}{c}{ \textbf{Out-of-Sample}}\\
&{\small \textbf{Accuracy}}&{\small \textbf{Recall}}&{\small \textbf{Precision}}&{\small \textbf{F1}}&{\small \textbf{PR AUC}}&{\small \textbf{ROC AUC}}&{\small \textbf{Accuracy}}&{\small \textbf{Recall}}&{\small \textbf{Precision}}&{\small \textbf{F1}}&{\small \textbf{PR AUC}}&{\small \textbf{ROC AUC}}\\\hline
logit& 85\% & 38\% & 72\% & 50\% &60\% &67\% & 83\%& 71\%& 57\%& 63\%& 67\% & 79\% \\ 
weighted logit & 79\% & 69\% &47\%  & 56\%  &60\%& 75\%& 80\%& 76\%& 50\%& 60\%&67\% & 78\%\\\hline
NN-Recall & 86\% &40\%  &77\%  & 53\% & 64\%&69\%& 86\%& 74\%& 62\%& {\color{red}\textbf{68\%}}& 67\% & 81\% \\
NN-Accuracy & 88\% &48\%  &85\%  & 61\% & 78\%&73\%& 88\%& 59\%& 76\%& 66\%& 69\% & 77\% \\
NN-F1 & 87\% &59\%  &68\%  & 63\% & 65\%&76\%& 86\%& 71\%& 64\%& 67\%& 71\% & 80\% \\ \hline
SMOTE-NN-Recall & 74\% &58\%  &79\%  & 67\% & 77\%&73\%& 62\%& {\color{red}\textbf{85\%}}& 33\%& 48\%& 63\% & 71\% \\
SMOTE-NN-Accuracy & 84\% &78\%  &84\%  & 81\% & 89\%&83\%& 88\%& 56\%& 76\%& 65\%& 70\% & 76\% \\
SMOTE-NN-F1 & 79\% &72\%  &79\%  & 83\% & 75\%&78\%& 86\%& 73\%& 62\%& 67\%& 68\% & 81\% \\
\hline
\end{tabular}}
\end{adjustwidth}
\end{center}
\caption{Results for \nth{1} Classification Framework}
\label{table:result1}
\end{table}
\FloatBarrier

In the second set of experiments, we add sentiment scores into the modeling framework. Table \ref{table:result2} and Table \ref{table:result3} illustrates the results obtained for \nth{2} Classification Framework and \nth{3} Classification Framework that are introduced in Section \ref{Methodology}. Similar to Table \ref{table:result1}, we compare our results against baseline models, and use Bayesian Optimization for hyperparameter tuning. To begin with, \texttt{LSTM Autoencoder} for sentiment scores is built separately, and the final autoencoder is obtained with hyperparameter tuning. Accordingly, we use 5 hidden units with \texttt{sigmoid} activation function. We then follow the framework in Figure \ref{plot:classification2} for the neural network structure. One hidden layer with 32 neurons and \texttt{selu} activation function is used for \texttt{Recall-NN} model under Focal Loss function. Similarly, one hidden layer with 64 neurons and \texttt{relu} activation function is used for \texttt{NN-Accuracy} model under Focal Loss function, and one hidden layer with 32 neurons and \texttt{elu} activation function is used for \texttt{NN-F1} model under F1 loss function. According to preliminary results, the best recall value is obtained for \texttt{NN-Recall} model by sacrificing overall accuracy. In order to improve the model performance, \texttt{SMOTE} algorithm is applied on the reduced dataset. \texttt{SMOTE-NN-Recall} model is built with two hidden layers with 32 neurons each. \texttt{Selu} and \texttt{elu} activation functions are used for hidden layers, and Tversky Loss, that is defined in Equation \ref{tversky_loss}, is used as the loss function. \texttt{SMOTE-NN-Accuracy} model uses two hidden layers with 32 and 8 neurons and \texttt{selu} activation function. Cross-entropy function is chosen as the loss function. \texttt{SMOTE-NN-F1} model uses two hidden layers with 32 and 16 neurons and \texttt{selu} activation function. As in previous framework, F1 loss is chosen as the loss function. The overall framework performance indicates that the results do not improve so much compared to the results in Table \ref{table:result1}. However, a broader range for the hyperparameters in hyperparameter tuning may increase the performance further. We will revise the results in the next review.

\begin{table}[htbp]
\vskip\baselineskip 
\begin{center}
\begin{adjustwidth}{-1.6cm}{0cm}
\scalebox{0.68}{
\begin{tabular}{ l|| c c c c c c|c c c c c c}\hline
\textbf{Methodology} & \multicolumn{6}{c|}{\textbf{In-Sample}}&\multicolumn{6}{c}{ \textbf{Out-of-Sample}}\\
&{\small \textbf{Accuracy}}&{\small \textbf{Recall}}&{\small \textbf{Precision}}&{\small \textbf{F1}}&{\small \textbf{PR AUC}}&{\small \textbf{ROC AUC}}&{\small \textbf{Accuracy}}&{\small \textbf{Recall}}&{\small \textbf{Precision}}&{\small \textbf{F1}}&{\small \textbf{PR AUC}}&{\small \textbf{ROC AUC}}\\\hline
logit& 85\% & 38\% & 72\% & 50\% &60\% &67\% & 83\%& 71\%& 57\%& 63\%& 67\% & 79\% \\ 
weighted logit & 79\% & 69\% &47\%  & 56\%  &60\%& 75\%& 80\%& 76\%& 50\%& 60\%&67\% & 78\%\\\hline
NN-Recall & 88\% &49\%  &79\%  & 60\% & 72\%&73\%& 67\%& {\color{red}\textbf{77\%}}& 35\%& 48\%& 55\% & 70\% \\
NN-Accuracy & 87\% &42\%  &80\%  & 56\% & 70\%&70\%& 88\%& 60\%& 75\%& 67\%& 69\% & 77\% \\
NN-F1 & 87\% &47\%  &78\%  & 58\% & 70\%&72\%& 87\%& 67\%& 67\%& 67\%& 71\% & 79\% \\ \hline
SMOTE-NN-Recall & 78\% &66\%  &80\%  & 72\% & 82\%&76\%& 83\%& 73\%& 56\%& 64\%& 69\% & 79\% \\
SMOTE-NN-Accuracy & 86\% &84\%  &84\%  & 84\% & 90\%&85\%& 87\%& 54\%& 77\%& 63\%& 72\% & 75\% \\
SMOTE-NN-F1 & 78\% &69\%  &80\%  & 74\% & 83\%&77\%& 87\%& 63\%& 70\%& 66\%& 70\% & 78\% \\
\hline
\end{tabular}}
\end{adjustwidth}
\end{center}
\caption{Results for \nth{2} Classification Framework}
\label{table:result2}
\end{table}
\FloatBarrier

\begin{table}[htbp]
\vskip\baselineskip 
\begin{center}
\begin{adjustwidth}{-1.6cm}{0cm}
\scalebox{0.68}{
\begin{tabular}{ l|| c c c c c c|c c c c c c}\hline
\textbf{Methodology} & \multicolumn{6}{c|}{\textbf{In-Sample}}&\multicolumn{6}{c}{ \textbf{Out-of-Sample}}\\
&{\small \textbf{Accuracy}}&{\small \textbf{Recall}}&{\small \textbf{Precision}}&{\small \textbf{F1}}&{\small \textbf{PR AUC}}&{\small \textbf{ROC AUC}}&{\small \textbf{Accuracy}}&{\small \textbf{Recall}}&{\small \textbf{Precision}}&{\small \textbf{F1}}&{\small \textbf{PR AUC}}&{\small \textbf{ROC AUC}}\\\hline
logit& 85\% & 38\% & 72\% & 50\% &60\% &67\% & 83\%& 71\%& 57\%& 63\%& 67\% & 79\% \\ 
weighted logit & 79\% & 69\% &47\%  & 56\%  &60\%& 75\%& 80\%& 76\%& 50\%& 60\%&67\% & 78\%\\\hline
NN-Recall & 84\% &56\%  &60\%  & 58\% & 61\%&73\%& 83\%& 71\%& 56\%& 63\%& 66\% & 78\% \\
NN-Accuracy & 86\% &43\%  &70\%  & 54\% & 62\%&69\%& 87\%& 62\%& 71\%& 66\%& 69\% & 78\% \\
NN-F1 & 86\% &43\%  &70\%  & 54\% & 62\%&69\%& 87\%& 62\%& 71\%& 66\%& 69\% & 78\% \\ \hline
SMOTE-NN-Recall & 72\% &71\%  &67\%  & 69\% & 70\%&72\%& 67\%& {\color{red} \textbf{81\%}}& 36\%& 49\%& 49\% & 72\% \\
SMOTE-NN-Accuracy & 81\% &75\%  &80\%  & 78\% & 84\%&80\%& 88\%& 62\%& 72\%& 67\%& 69\% & 78\% \\
SMOTE-NN-F1 & 81\% &75\%  &80\%  & 78\% & 84\%&80\%& 88\%& 62\%& 72\%& 67\%& 69\% & 78\% \\
\hline
\end{tabular}}
\end{adjustwidth}
\end{center}
\caption{Results for \nth{3} Classification Framework}
\label{table:result3}
\end{table}

\FloatBarrier
Table \ref{table:result3} shows the results obtained for \nth{3} Classification Framework. As illustrated in Figure \ref{plot:classification3}, this model directly takes sentiment scores into the model as input. An \texttt{LSTM} layer then reads these sentiment scores and outputs the final internal state. At the same time, the reduced dataset of numerical and categorical variables are used as an input to a vanilla \texttt{FFNN}, and its output is concatenated together with \texttt{LSTM} output. The merged dataset is then fed into a \texttt{FFNN} and the result provides the prediction of the deal status. Compared to \nth{2} Classification Framework, all components of the model are trained at the same time. Again, a similar procedure is used for obtaining results. \texttt{NN-Recall} model uses two hidden layers with 4 and 8 neurons. \texttt{Selu} activation function is used in both layers for nonlinearity, and F1 Loss is chosen as the loss function. \texttt{NN-Accuracy} and \texttt{NN-F1} models use the same set of hyperparameters. They both use two hidden layers with 4 and 16 neurons, respectively. \texttt{Elu} activation function is used in both layers together with Focal loss function. \texttt{NN Recall} model results are comparable with the results of benchmark models. When \texttt{SMOTE} is applied for oversampling, \texttt{SMOTE-NN-Recall} model attains the highest recall value by decreasing overall accuracy performance. \texttt{SMOTE-NN-Recall} model consists of two hidden layers with 4 and 16 neurons, respectively. Nonlinearity is introduced into the model using \texttt{selu} activation function, and F1 loss is chosen as the loss function. The best overall accuracy and f1 scores are obtained by using the same model architecture. The model consists of two hidden layers with 64 neurons each. \texttt{Selu} and \texttt{relu} activation functions are used, and Tversky loss function is chosen as the objective function. Our results indicate that the results are compatible with the results obtained from \nth{2} Classification Framework. The best overall performance is attained with \texttt{NN-Recall} model which is very similar to \texttt{SMOTE-NN-Recall} model results from \nth{2} Classification Framework. A more thorough hyperparameter tuning will be applied for the next revision.

\section{Conclusion and Future Direction}\label{Conclusion}
In this study, we utilize a large number of numeric and categorical features and a temporal dataset of sentiment scores to predict the status of the merger \& acquisition deals. To make the data denser, we apply \texttt{PCA} on numeric variables and \texttt{MCA} on categorical variables. We further employ \texttt{LSTM Autoencoder} on sentiment scores in order to be able to use them in a vanilla \texttt{FFNN} setting. Our preliminary results indicate that our model sets a better balance between overall accuracy and recall compared to baseline models such as logit model. We obtain higher accuracy levels for comparable recall levels. Our results also show that use of sentiment scores in classification framework does not improve the performance further compared to \nth{1} Classification Framework. We expect that a broader hyperparameter range may increase the performance of \nth{2} and \nth{3} Classification Frameworks. To the best of our knowledge, dimensionality reduction techniques have not been used in takeover success prediction problem. Furthermore, sentiment scores are not used for takeover success prediction in the literature before. We build two preliminary frameworks to introduce sentiment scores into takeover success prediction problem. \nth{2} Classification Framework consists of \texttt{LSTM Autoencoder}, and we train \texttt{LSTM Autoencoder} earlier to provide input into \texttt{FFNN} model. However, \texttt{LSTM Autoencoder} adds an additional noise into the input. Therefore, the \nth{3} Classification Framework integrates the sentiment scores directly into the model, and overall training time is faster than the \nth{2} Classification Framework yet the results are similar. A more thorough approach for using sentiment scores will be added later as a future work. In this paper, we introduce our overall methodology for Post-Announcement part of the model. We apply a similar methodology on Pre-Announcement model, and the results for Pre-Announcement will be added in the next version of our study. 

\clearpage

\printbibliography


\end{document}